\input{aipcheck}

\documentclass[
    ,final            
  ]
  {aipproc}

\layoutstyle{6x9}

\renewcommand\XFMtitleblock{
  \XFMtitle
  \let\XFMoldpar\par
  \def\par{\XFMoldpar\def\par{\space
             on behalf of the {\it Fermi}-LAT Collaboration\XFMoldpar}}%
   \XFMauthors
   \let\par\XFMoldpar
   \XFMaddresses
   \XFMabstract
   \vspace{5pt}%
   \XFMkeywords
   \XFMclassification
 }

\begin{document}

\title{To be or not to be a blazar. The case of the Narrow-Line Seyfert 1 SBS 0846$+$513}


\author{F. D'Ammando}{address={Dip. di Fisica, Universit\'a degli Studi di Perugia and INFN, Via A. Pascoli, I-06123 Perugia, Italy }}

\author{M. Orienti}{address={Dip. di Astronomia, Universit\'a di Bologna, Via Ranzani 1, I-40127 Bologna, Italy \\ INAF - Istituto di Radioastronomia,  Via Gobetti 101, I-40129 Bologna, Italy}}

\author{J. Finke}{address={U.S. Naval Research Laboratory, Code 7653, 4555 Overlook Ave. SW, Washington, DC 20375-5352, USA}} 

\begin{abstract}
The presence of a relativistic jet in some radio-loud Narrow-Line Seyfert 1s (NLSy1) galaxies, first suggested by their variable radio emission and the
flat radio spectra, is now confirmed by the {\em Fermi}-LAT detection of five
NLSy1s in $\gamma$ rays. In particular, a strong $\gamma$-ray flare from SBS
0846$+$513 was observed in 2011 June by {\em Fermi}-LAT reaching a $\gamma$-ray luminosity (0.1--300
GeV) of $\sim$10$^{48}$ erg s$^{-1}$, comparable to that of bright flat spectrum radio quasars. Apparent superluminal velocity in the jet was inferred from 2011--2012 VLBA images, suggesting the presence of a highly relativistic jet.
Both the power released by this object during the flaring activity and the
apparent superluminal velocity are strong indicators of the presence of a
relativistic jet as powerful as those in blazars. In addition, variability and
spectral properties in radio and $\gamma$-ray bands indicate a blazar-like
behaviour, suggesting that, except for some distinct optical characteristics,
SBS 0846$+$513 could be considered as a young blazar at the low end of the blazar's black hole mass distribution.
\end{abstract}

\maketitle


\section{Narrow-Line Seyfert 1 galaxies and {\em Fermi}-LAT}

Relativistic jets are
the most extreme expression of the power that can be generated by a
super-massive black hole (SMBH) in the center of an AGN.  These objects
have a total bolometric luminosity of up to 10$^{49-50}$ erg s$^{-1}$
\citep[e.g.][]{ackermann10, bonnoli11}, with a large fraction of the power emitted at high energies. Before the launch of
the {\em Fermi} satellite only two classes of AGN were known to
generate relativistic jets and to emit in the $\gamma$-ray energy
range: blazars and radio galaxies, both hosted in giant elliptical
galaxies \citep{blandford78}. The first two years of observations
by {\em Fermi}-LAT confirmed that these two populations represent the
majority of the identified sources in the extragalactic $\gamma$-ray sky
\citep{abdo10a, nolan12}. However the discovery of variable $\gamma$-ray
emission from four radio-loud NLSy1 galaxies revealed the
presence of a possible third class of AGN with relativistic jets
\citep{abdo09}.  

The NLSy1s are a class of AGN identified by \citep{osterbrock85} and characterized by
the following optical properties: narrow permitted lines (FWHM (H$\beta$) $<$
2000 km s$^{-1}$), [OIII]/H$\beta$ $<$ 3, and a Fe II bump \citep[for a review
see][]{pogge00}. 
They have smaller central black hole masses (i.e.~10$^6$-10$^8$
M$_\odot$) than those in blazars and radio galaxies, and higher accretion
rates close to or above the Eddington limit \citep{yuan08}. These sources
are generally radio-quiet, with only a small fraction ($<$7\%) classified as
radio-loud \citep{komossa06}, and objects with high values of radio-loudness ($R > 100$)
are even more sparse ($\sim$2.5\%). The strong radio emission and the flat radio spectrum, together with variability studies, suggest the presence of
a relativistic jet in some radio-loud NLSy1s.  This has been confirmed by the
{\it Fermi}-LAT detection of $\gamma$-ray emission from some of them. By considering that NLSy1s are
usually hosted in spiral galaxies \citep[e.g.][]{deo06} the presence
of a relativistic jet is in contrast to the paradigm that the formation
of relativistic jets could happen only in elliptical galaxies \citep[see e.g.][]{marscher10}, giving an indication that relativistic jets can
form and develop independently of their host galaxies.

\section{Fermi-LAT observations}

SBS 0846$+$513 was not in the first and second Fermi-LAT (1FGL and 2FGL)
catalogues, indicating that the source was not detected with Test Statistic (TS)
$>$ 25 in either one year or two
years of {\em Fermi} observations \citep{abdo10a,nolan12}. Integrating over
the first two years of {\em Fermi} operation the likelihood fit yielded a TS = 14, with a
2-$\sigma$ upper limit of 8.5$\times$10$^{-9}$ photons cm$^{-2}$ s$^{-1}$ in the 0.1--300 GeV energy
range, assuming a photon index $\Gamma=2.3$. On the contrary, the likelihood fit with a power-law model
to the data integrated over the third year of {\em Fermi} operation
(2010 August 4 -- 2011 August 4; MJD 55412--55777) in the 0.1--300 GeV
energy range results in a TS = 653, with an integrated average flux of
(6.7 $\pm$ 0.5) $\times$10$^{-8}$ ph cm$^{-2}$ s$^{-1}$ and a
photon index of $\Gamma$ = 2.23 $\pm$ 0.05.

\begin{figure}
 \begin{minipage}[b]{5cm}
   \centering
   \includegraphics[width=6cm]{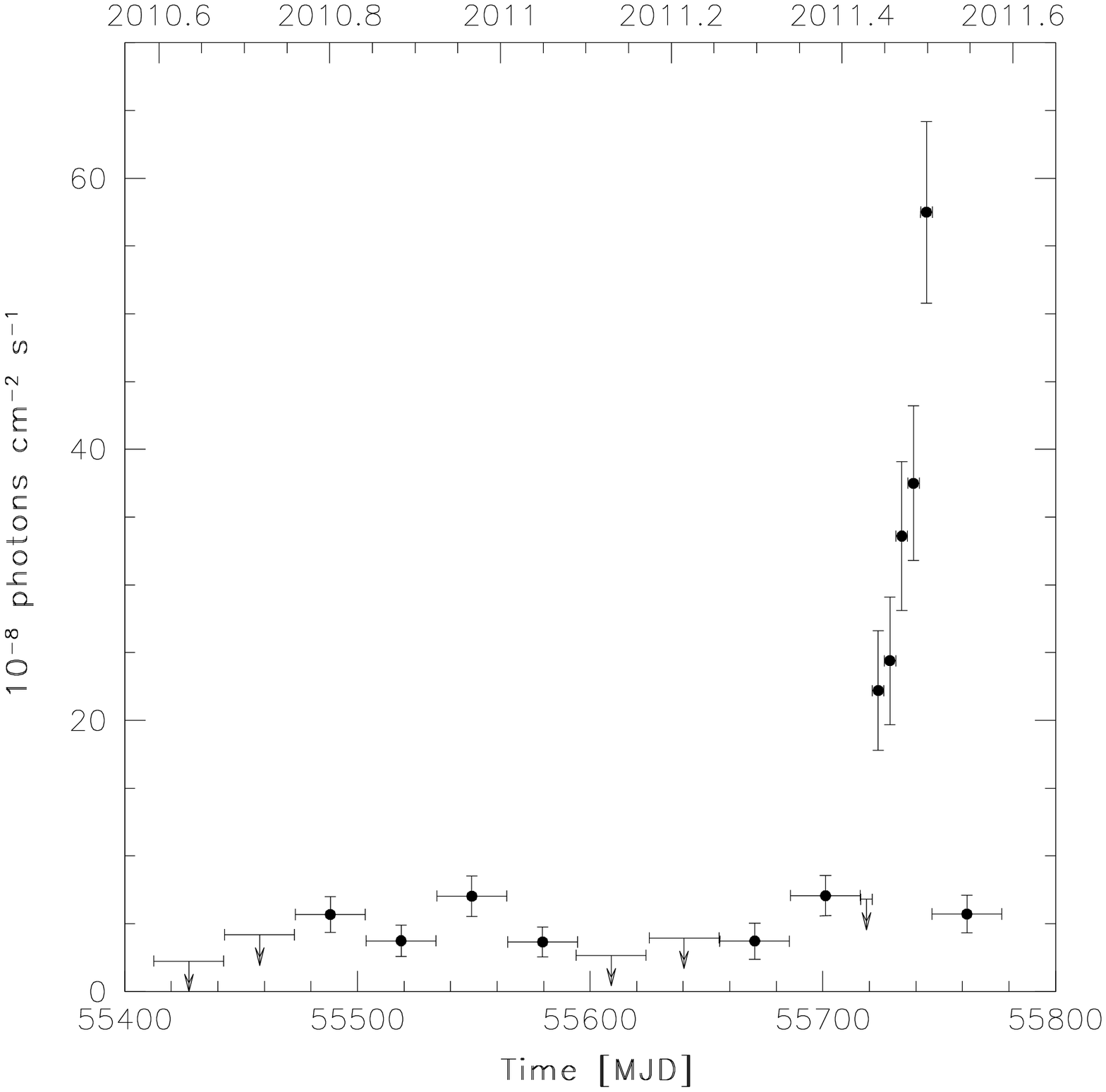}
   \caption{{\it Left panel:} 1-month and 5-day integrated flux (E $>$ 100 MeV) light curve of SBS\,0846$+$513
  obtained from 2010 August 4 to 2011 August 4. Arrows refer to 2-$\sigma$
  upper limits on the source flux. {\it Right panel:} VLBA image at 15.3 GHz of SBS 0846+513. On the image we provide the
peak flux density, in mJy/beam, and the first contour intensity (f.c., in mJy/beam) that corresponds to three times the noise measured on the
image plane. Contour levels increase by a factor of 2. The beam is
plotted on the bottom left corner of the image. The vectors superimposed on the total intensity contours show the position angle of the E vector, where 1 mas length corresponds to 2.9 mJy/beam.}\label{Fig1}
 \end{minipage}
 \ \hspace{10mm} \hspace{5mm} \
 \begin{minipage}[b]{5cm}
  \centering
   \includegraphics[width=4.5cm]{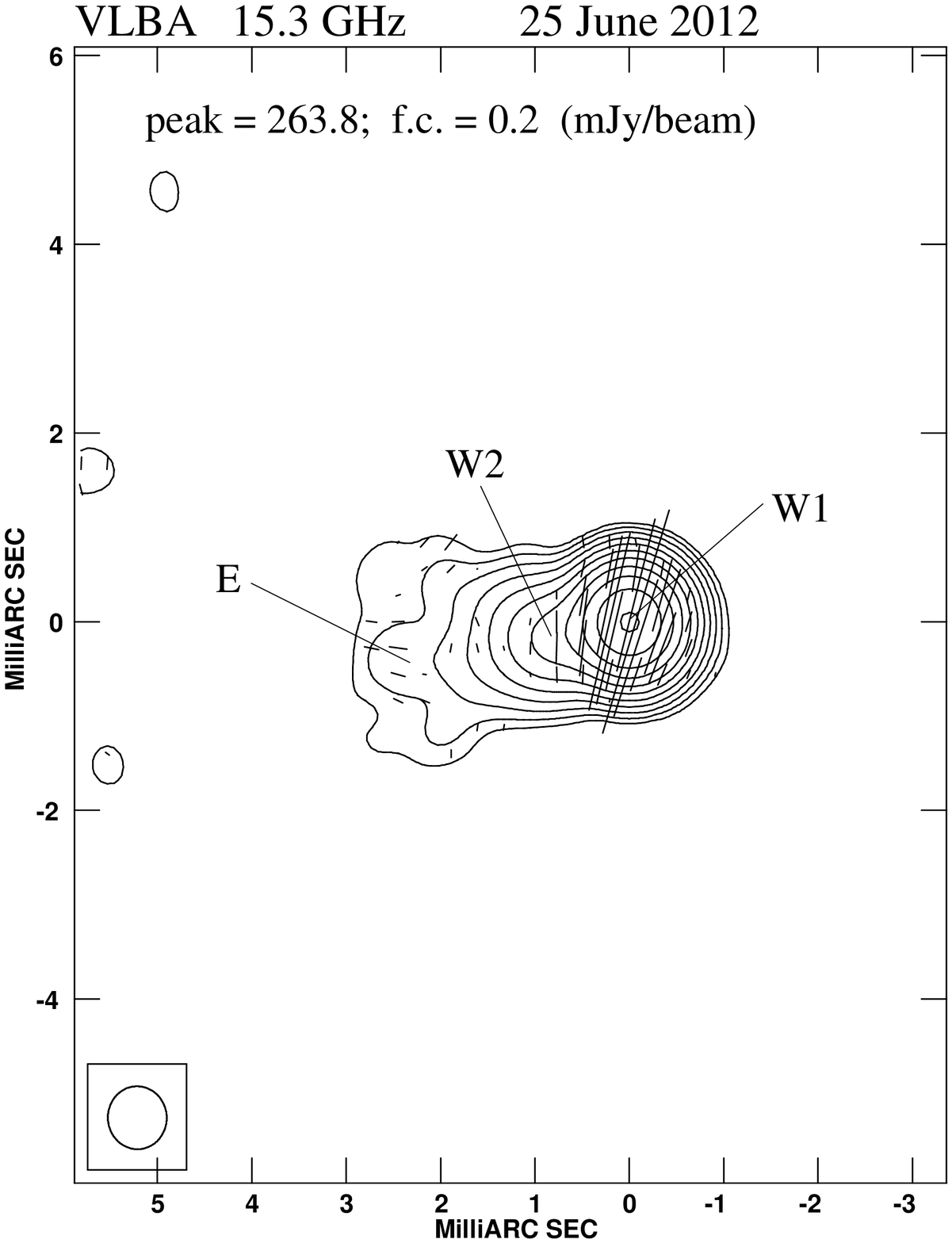}
 \end{minipage}
\end{figure}

Figure~\ref{Fig1} (left panel) shows the $\gamma$-ray light curve of the third year of {\em Fermi}
observations built using 1-month time bins. If TS $<$ 10 the value of the
flux was replaced by the 2-$\sigma$ upper limits. A clear increase of the
flux was observed in the period 2011 June 4--July 4, for which 5-day time
bins were used in Fig.~\ref{Fig1}. Considering the high activity of the source we extracted a spectrum over
that period, obtaining a photon index of $\Gamma$ = 1.98 $\pm$ 0.05 and a flux of
(24.4 $\pm$ 2.1) $\times$10$^{-8}$ ph cm$^{-2}$ s$^{-1}$. The spectral
evolution during the flaring activity in 2011 June observed in $\gamma$ rays from SBS 0846$+$513 is a common behaviour in bright
FSRQs and low-synchrotron-peaked BL Lacs detected by {\em Fermi} \citep{abdo10b}, with a change in photon index $<$ 0.2--0.3.
The peak of the emission was observed on 2011 July 1, with a flux of (98 $\pm$ 19) $\times$10$^{-8}$
ph cm$^{-2}$ s$^{-1}$ in the 0.1--300 GeV energy range, corresponding to
an isotropic $\gamma$-ray luminosity of $\sim$1.1$\times$10$^{48}$ erg
s$^{-1}$. Further details about the LAT data analysis and results are given in \citep{dammando12}.

\section{The radio morphology and variability}

When imaged with the high spatial resolution of the VLBA SBS 0846$+$513 is resolved in two components with a core-jet structure, 
as also pointed out in a previous work by \citep{taylor05}. At 15 GHz the jet
structure (component E in Fig.~\ref{Fig1}, right panel) shows an extended low-surface brightness structure with a steep
spectrum. On the other hand the core (component W) is resolved into two compact
components (labelled W1 and W2 in Fig.~\ref{Fig1}, right panel). To investigate a possible proper motion of the jet, 
we compared the separation between W1, considered the core region, and
W2, assumed to be a knot in the jet, at the four MOJAVE observing epochs
available up to now. We model-fitted the visibility data using gaussian components of the
four-epoch data by means of the model-fitting option in
\textsc{difmap}. From this comparison we found that W1 and W2 are separating with an apparent velocity of
(0.32 $\pm$ 0.04) mas/yr, which corresponds to (10.9 $\pm$ 1.4)$c$. This apparent
superluminal velocity suggests the presence of boosting effect. 

Before the $\gamma$-ray flaring episode, the simultaneous
multifrequency observations performed by Effelsberg showed a flat
radio spectrum up to 32 GHz. After the flare, the spectral shape observed changed, becoming convex. The spectral variability was also accompanied
by variations in the radio flux density, as observed by OVRO and Medicina
telescopes, indicating a blazar-like behaviour. Further details are given in \citep{dammando12}.

\section{Discussion}

After PMN J0948$+$0022 \citep{foschini_etal11}, SBS 0846$+$513 is the second NLSy1 observed
to generate such a high power in $\gamma$ rays. This could be an indication that all
the radio-loud NLSy1s are able to host relativistic jets as powerful as
those in blazars, despite the lower BH mass; alternatively some NLSy1s
could have peculiar characteristics allowing the development of these
relativistic jets. The mechanism at work for producing a relativistic
jet is not clear, and the physical parameters that drive the jet formation
is still under debate. One fundamental parameter could be the black
hole mass, with only large masses allowing for the efficient formation of a
relativistic jet. 
The large radio loudness of SBS 0846$+$513 could challenge this idea if the
black hole mass (between 8.2$\times10^6\ M_\odot$ and 5.2$\times10^7\ M_\odot$) estimated by \citep{zhou05} is confirmed. According
to the ``modified spin paradigm'' discussed in \citep{sikora07}, another
fundamental parameter for the efficiency of a relativistic jet production should be the BH spin, with SMBHs in elliptical galaxies having on average
much larger spins than SMBHs in spiral galaxies. 

In this context the discovery of relativistic jets in a class of AGN
usually hosted by spiral galaxies, such as the NLSy1s, was a great
surprise.
We note that BH masses of radio-loud NLSy1s are generally
larger with respect to the entire sample of NLSy1s \citep{komossa06, yuan08}, even if still
small when compared to radio-loud quasars. The larger BH masses of
radio-loud NLSy1s with respect to radio-quiet NLSy1s could be related to prolonged accretion episodes that
can spin-up the BHs. The small fraction of radio-loud NLSy1s with
respect to radio-loud quasars could be an indication that in the
former the high accretion usually does not last sufficiently long to
significantly spin-up the BH \citep{sikora09}. To conclude, SBS 0846$+$513 shows all the characteristics of the
blazar phenomenon and could be a low mass (and possible younger) version of blazar. 


\begin{theacknowledgments}
The {\em Fermi} LAT Collaboration acknowledges support from a number of agencies
and institutes for both development and the operation of the LAT as well as scientific data analysis. These include NASA and DOE in the United States, CEA/Irfu and
IN2P3/CNRS in France, ASI and INFN in Italy, MEXT, KEK, and JAXA in Japan,
and the K. A. Wallenberg Foundation, the Swedish Research Council and the National
Space Board in Sweden. Additional support from INAF in Italy and CNES in France for
science analysis during the operations phase is also gratefully acknowledged. This research has made use of
data from the MOJAVE database that is maintained by the MOJAVE team
(Lister et al. 2009, AJ, 137, 3718). 
\end{theacknowledgments}



\bibliographystyle{aipprocl} 




\end{document}